\documentclass[aps,pre,twocolumn,superscriptaddress,showpacs,nofootinbib]{revtex4-1}
\usepackage{graphicx}
\usepackage{bbold}
\usepackage{amsmath}
\usepackage{color}
\usepackage{amssymb}
\usepackage{hyperref}
\usepackage[space]{grffile}
\usepackage[caption=false]{subfig}
\usepackage[titletoc, title, toc]{appendix}
\usepackage{etoolbox}
\usepackage{multirow}
\usepackage[export]{adjustbox}
\usepackage{tikz}
\usetikzlibrary{arrows.meta}

\newcommand{\lt}{l_{\rm TPB}}
\newcommand{\Lt}{L_{\rm TPB}}
\newcommand{\hy}{\! \operatorname{-} \!}

\begin{document}

\title{Theory-based design of sintered granular composites \\ triples three-phase boundary in fuel cells}

\author{Shahar Amitai}
\email{s.amitai13@imperial.ac.uk}
\affiliation{Imperial College London, London SW7 2BP, UK}
\author{Antonio Bertei}
\affiliation{Imperial College London, London SW7 2BP, UK}
\author{Raphael Blumenfeld}
\affiliation{Imperial College London, London SW7 2BP, UK}
\affiliation{Cavendish Laboratory, Cambridge CB3 0HE, UK}

\date{\today}

\begin{abstract}

Solid-oxide fuel cells produce electric current from energy released by a spontaneous electrochemical reaction. The efficiency of these devices depends crucially on the microstructure of their electrodes and, in particular, on the three-phase boundary (TPB) length, along which the energy-producing reaction occurs. We present a systematic maximisation of the TPB length as a function of four readily-controllable microstructural parameters, for any given mean hydraulic radius, which is a conventional measure
of the permeability to gas flow. We identify the maximising parameters and show that the TPB length can be increased by a factor of over 300\% compared to current common practices. We support this result by calculating the TPB of several numerically simulated structures. We also compare four models for a single intergranular contact in the sintered electrode and show that the model commonly used in the literature is oversimplified and unphysical. We then propose two alternatives.

\end{abstract}

\maketitle

\section{Introduction} \label{sec:introduction}

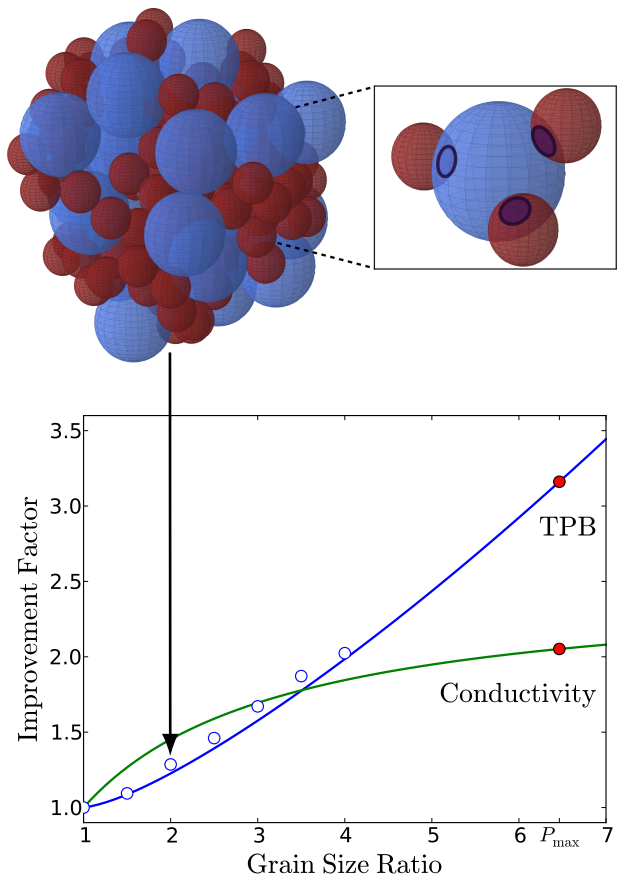
\begin{figure}[h]
\begin{tikzpicture}[remember picture]
  \node[anchor=south west, inner sep=0] (imageA) {\includegraphics[width=0.47\textwidth]{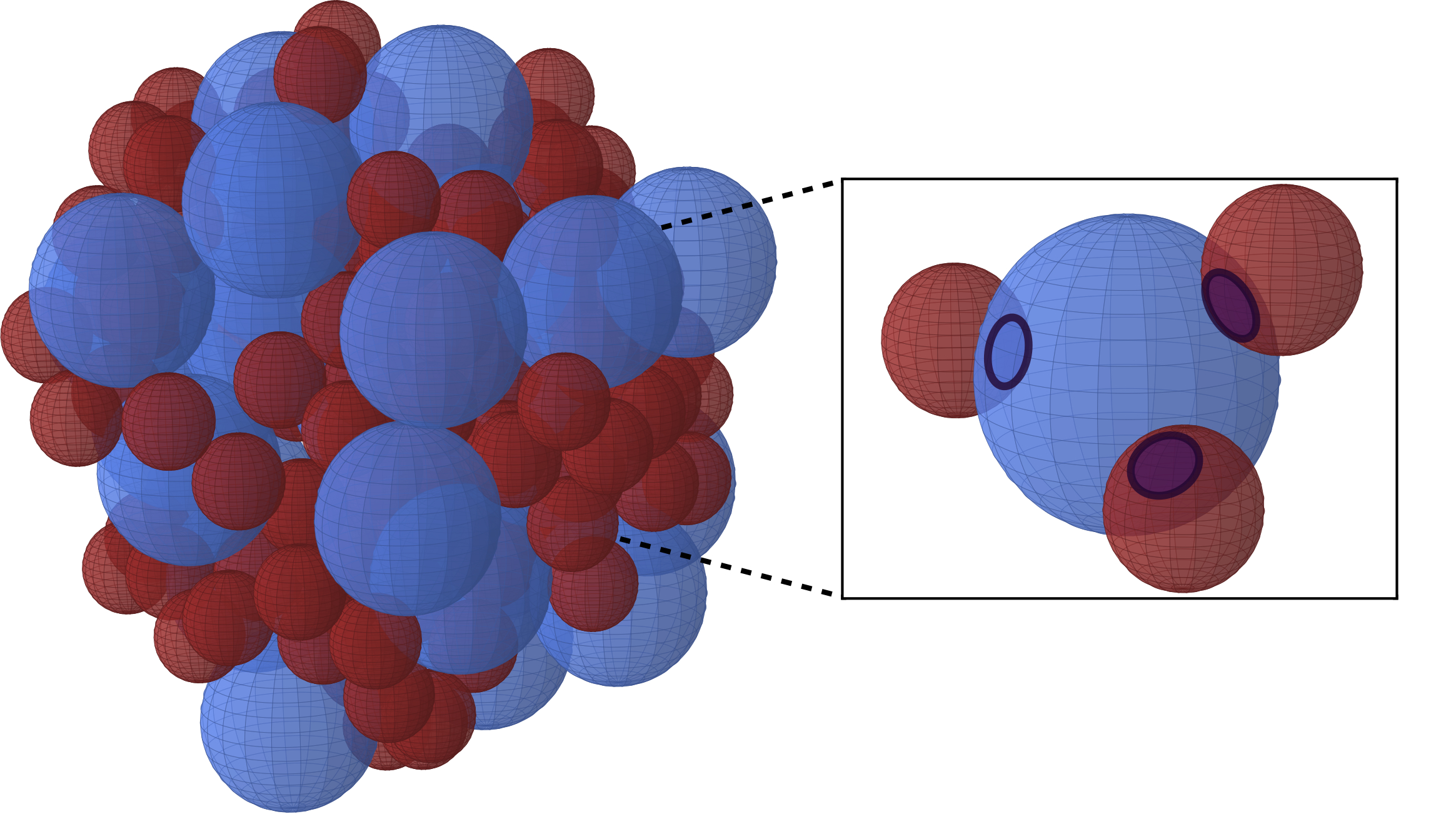}};
  \node (A) at (2.165,0.25) {};
\end{tikzpicture}
\begin{tikzpicture}[remember picture]
  \node[anchor=south west,inner sep=0] (imageB) {\includegraphics[width=0.5\textwidth]{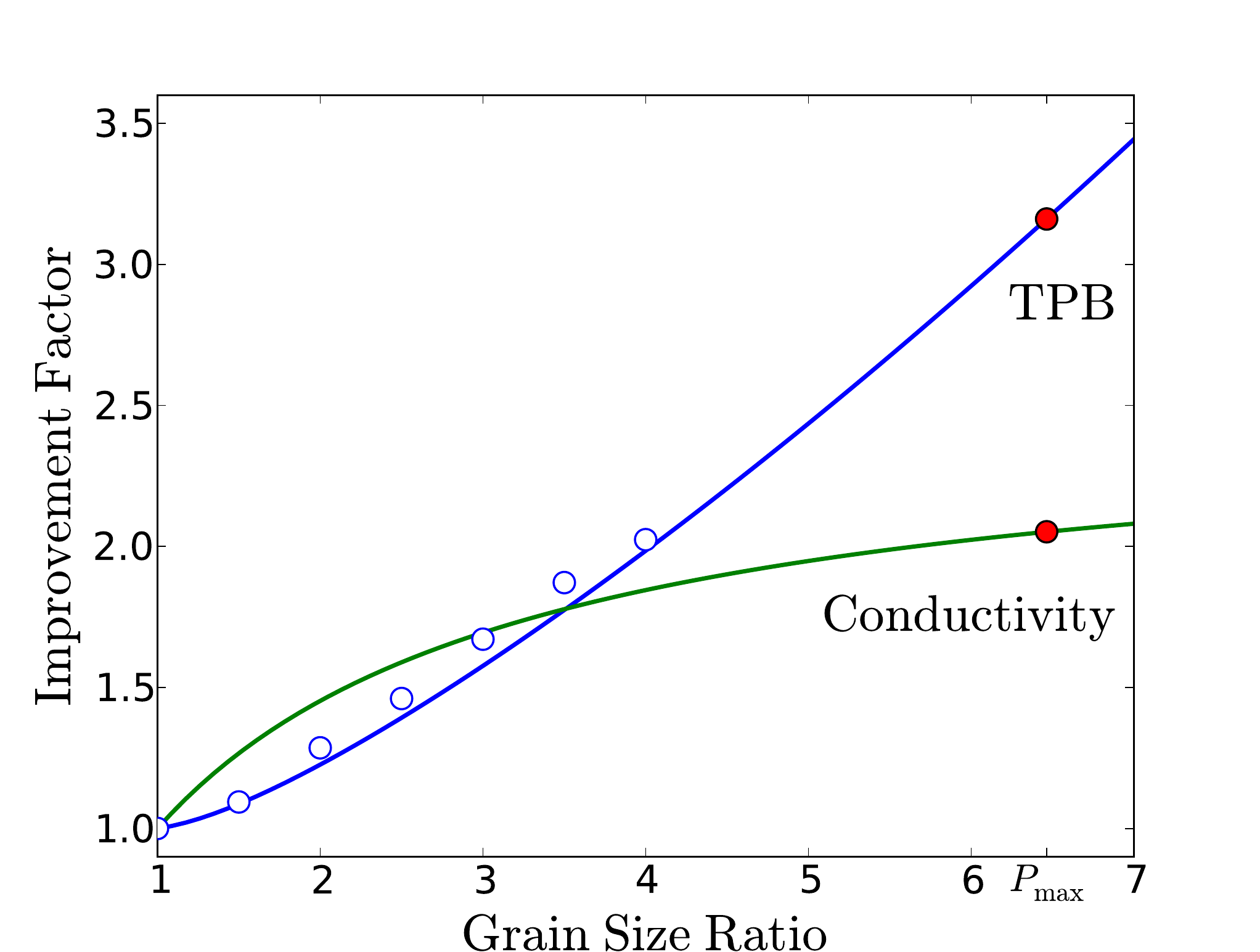}};
  \node (B) at (2.28, 1.45) {};
\end{tikzpicture}
\begin{tikzpicture}[remember picture,overlay]
  \draw[-{Latex[length=3mm]}, line width=1pt] (A) -- (B);
\end{tikzpicture}
\caption{Top: A simulated composite electrode, made of a sintered powder of ion-conducting (blue) grains of radius $r_1$ and electron-conducting (red) grains of radius $r_2 = r_1 / P$ with $P = 2$. The `zoom-in' window shows three inter-phase contact discs and their respective TPBs as dark circular lines. Bottom: our predicted improvement factor for TPB density (blue line) and conductivity (green line) as functions of $P$, compared to the common design of $P = 1$, for the same mean hydraulic radius. The red dots are respective maximal improvements, obtained for $P_{\rm max}$. The blue circles are measured TPB improvements for simulated systems. These match our prediction to an accuracy of 6\%.}
\label{fig:intro}
\end{figure}

Electrodes of solid-oxide fuel cells (SOFCs) commonly comprise a porous composite material, made by randomly packing and then sintering a mixture of electron conducting and ion conducting powders. The interfaces between particles of the two phases (contacts) that are exposed to the pore space form a collection of closed lines, sketched in Fig. \ref{fig:intro}, called the triple-phase boundary (TPB). It is along the TPB that an energy producing electrochemical redox reaction takes place \cite{SinKen03}. The efficiency of the SOFC depends sensitively on its electrode microstructure and, in particular, on the spatial distribution and total length of the TPB per unit volume, $\Lt$. This quantity, which is a good indicator for the current that the fuel cell can produce, depends on the mean TPB length of a single inter-phase contact and the number of such contacts per unit volume.

We specialise the analysis to each powder consisting of spherical grains of a specific size, which leaves the microstructure of the porous medium depending on four parameters:
(a) the radii of the two types of grains, $r_1$ and $r_2$;
(b) the relative volume fractions of the powders, parametrised by $\psi_1$ -- the fraction of the total solid volume occupied by type-1 grains, with $\psi_2 = 1 - \psi_1$;
(c) the porosity, $\phi$.
While $\Lt$ was studied analytically \cite{Ja08, Chen09, Zha11, Ber11Perc}, by simulations \cite{Sch07, Gol08, Ken09, San10, Ge12, Ber12} and experimentally \cite{Wil06, Sm09, Faes09, She10}, there is no systematic study of the dependence of $\Lt$, or its maximum, on the above four parameters. Nor is there a fair comparison between different models of the contact geometry between grains of different sizes. Here, we address both these issues: compare and assess different models of a single contact geometry and find a set $\{r_1, r_2, \psi_1, \phi \}$ that maximises $\Lt$.

Such maximisation cannot be unconstrained: $\Lt$ can be increased indefinitely by simply reducing both grain sizes for given $\psi_1$ and $\phi$. For example, reducing all radii by a factor $\gamma < 1$ increases the number of contacts by $1/\gamma^3$ while the TPB of every single contact decreases as $\gamma$, resulting in an overall increase of $\Lt$ as $1/\gamma^2$. However, the finer the powders the less permeable is the sintered electrode to flow of the gaseous reactants and products. The permeability can be parameterised by the mean hydraulic radius, $R_h$, defined below, which increases monotonically with both grain sizes. Too small $R_h$ restricts gas flow, reducing the efficiency of the SOFC \cite{Ber13}. Therefore, the following analyses is carried out at fixed $R_h$.

Our main result is summarised in Fig. \ref{fig:intro}: for any given value of $R_h$, the TPB can be increased systematically by increasing the grain size ratio $P \equiv r_1 / r_2$. The longest TPB obtained for a mix of $P_{\rm max} \simeq 6.46$. Fig. \ref{fig:intro} also shows that not only is $\Lt$ 3.16 times longer for this mixture but also the effective conductivity of the solid phase is twice as high. Using a scaling argument, we show below that these conclusions hold for any value of $R_h$.

A summary of the maximisation process of $\Lt$ in the four-parameter space $\{r_1, r_2, \psi_1, \phi \}$ is as follows. Taking $r_1 \ge r_2 \ge 0$ and since $0 < \psi_1, \phi < 1$, any choice of these parameters describes a valid system. We first fix $R_h$ and $\phi$ to the reasonable values $R_h = 54$nm (justified in section \ref{sec:volume_fractions}) and $\phi = 0.36$ (the porosity of the maximally random jammed state \cite{Ber60}), but show that the specific choice of $R_h$ and $\phi$ does not affect the results. We then use the value of $R_h$ to calculate $\psi_1$ for each pair of values $(r_1, r_2)$. From knowledge of $r_1$, $r_2$ and $\psi_1$ we calculate $\Lt$. We then compare $\Lt$ of a monodisperse mixture with any bidisperse one of equal $R_h$ and $\phi$, and determine the size combination that maximises $\Lt$.

The rest of this paper is organised as follows. 
In section \ref{sec:volume_fractions}, we determine $R_h$ and then $\psi_1$ for any combination $(r_1, r_2)$. 
In sections \ref{sec:single_tpb}, \ref{sec:number_of_contacts} and \ref{sec:active_fraction}, we calculate the quantities on which $\Lt$ depends: the mean TPB length of a single contact, the number of contacts per unit volume, and the percolating fraction of each phase. We then assemble the results to obtain $\Lt$ and identify the grain sizes that maximise it. 
In section \ref{sec:other_features}, we discuss the ideal size combination and show that it also leads to a higher effective conductivity. 
In section \ref{sec:recipe}, we discuss the scaling properties of our solution and use our results to prescribe an optimal design of composite electrodes. 
In section \ref{sec:numerical} we present numerical simulations that support our predictions. 
We conclude with a summary and general discussion in section \ref{sec:conclusions}.

\section{The volume fraction and the mean hydraulic radius} \label{sec:volume_fractions}

Many parameters and constraints affect overall electrode performance, complicating equal-footing comparison of different structures. We regard structures as fairly compared when they have the same packing fraction, $\phi$, and hydraulic radius, $R_h$. The latter is a measure of the electrode's permeability to gas flow and it is commonly defined as the ratio between the total pore volume and  solid surface area \cite{Ber11Perc},
\begin{align} \label{eq:r_h_orig}
R_h = \frac{1 - \frac{4 \pi}{3} \left( n_1 r_1^3 + n_2 r_2^3 \right)}{4 \pi \left( n_1 r_1^2 + n_2 r_2^2 \right)} \ ,
\end{align}
where $n_i$ ($i = 1, 2$) is the number of grains of phase $i$ per unit volume. The solid volume fraction is $(1 - \phi) = \frac{4 \pi}{3} \left( n_1 r_1^3 + n_2 r_2^3 \right)$, where $\phi$ is the porosity. It is convenient to express $R_h$ in terms of the number fractions, $\xi_i \equiv n_i / \left(n_1 + n_2\right)$,
\begin{align} \label{eq:r_h}
R_h = \frac{\phi \left( \xi_1 r_1^3 + \xi_2 r_2^3 \right)}{3 (1 - \phi) \left( \xi_1 r_1^2 + \xi_2 r_2^2 \right)} \ ,
\end{align}
in terms of which we wish to calculate the volume fraction, $\psi_1$, for any value of $r_2$ and $r_1 \ge r_2$. In the equivalent monodisperse system $r_1 = r_2 = r_0$ and eq. (\ref{eq:r_h}) reduces to $R_h = \frac{\phi r_0}{3 (1 - \phi)}$. This gives the grain size of the equivalent monodisperse system,
\begin{align} \label{eq:r_h_simple}
r_0 = \frac{\xi_1 r_1^3 + \xi_2 r_2^3}{\xi_1 r_1^2 + \xi_2 r_2^2} \qquad (r_2 \le r_0 \le r_1) \ ,
\end{align}
and makes it possible to express the number fractions and solid fractions in terms of $r_0$, $r_1$ and $r_2$ as follows. Using eq. (\ref{eq:r_h_simple}) and $\xi_1 + \xi_2 = 1$, we obtain $\xi_1$ (and $\xi_2$):
\begin{align}
\xi_1 = \frac{r_2^2 (r_0 - r_2)}{r_1^3 - r_2^3 - r_0 (r_1^2 - r_2^2)} \ ,
\end{align}
from which one determines the volume fractions:
\begin{align} \label{eq:number_to_volume_fraction}
\psi_{i} = \frac{\xi_{i}r_i^3}{\xi_{1}r_1^3 + \xi_{2} r_2^3} \qquad ; \qquad i = 1,2 \ .
\end{align}

Not any combination of $r_1$ and $r_2$ is useful: if the ratio $r_1/r_2\equiv P$ is too large the small particles fall through the interstices and the phases segregate. This bounds $P$ to below $P_{\rm max} \simeq 6.46$ \cite{Kuo95}.

To determine $R_h$, we resort to another measure of the flow rate -- the Knudsen number, $K_n = \lambda / 4 R_h$, with $\lambda$ the gas mean free path. For typical SOFC operating conditions, $800^\circ C$ and $1$atm, the hydrogen mean free path is $\lambda = 0.432\mu$m and, using an accepted optimal value in the literature, $K_n\approx 2$ \cite{Ber13}, we obtain $R_h \approx 54$nm. This value corresponds to $r_0 \simeq 0.3\mu$m. In the following, we use this value for $r_0$ and all grain radii satisfy $r_2 \le r_0 \le r_1$. Nevertheless, we emphasise that the following analysis is general in that it applies to any choice of mean hydraulic radius, $R_h \to \gamma R_h$, by scaling $r_i \to \gamma r_i$ ($i = 0,1,2$) and $\Lt \to \Lt / \gamma^2$.

\begin{figure}[h]
  \includegraphics[width=0.47\textwidth]{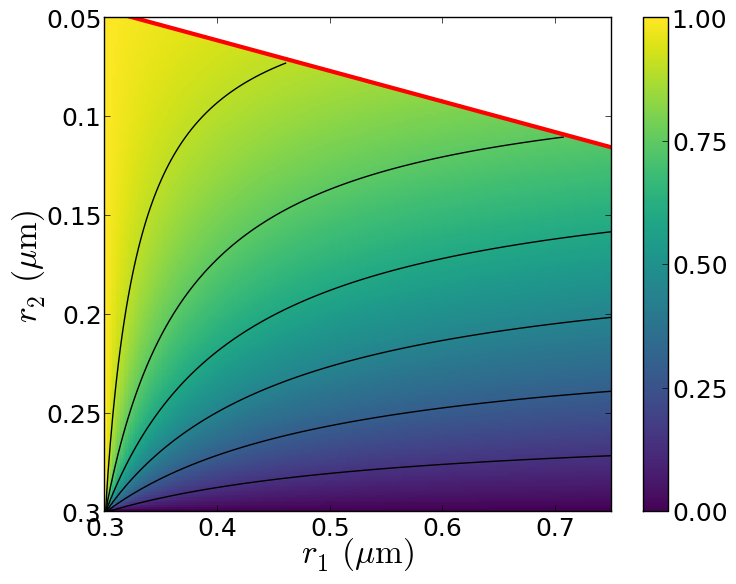}
  \caption{The volume fraction, $\psi_1$, as a function of the two grain radii, $r_1 \ge r_0\ge r_2$. In black are equal $\psi_1$ contour lines and the top thick line is the maximum-size ratio limit, $r_1 / r_2 = P_{\rm max}$. Note that the ordinate is labeled in decreasing values of $r_2$.}
  \label{fig:volume_fraction}
\end{figure}

In Fig. \ref{fig:volume_fraction} we show $\psi_1$, as a function of $r_1$ and $r_2$, for $R_h = 54$nm. The figure shows that the region of high (low) $\psi_1$ corresponds to values of $r_1$ ($r_2$) close to $r_0$.

\section{A single contact TPB} \label{sec:single_tpb}

The overall TPB length increases linearly with the mean TPB length of a single contact, $\lt$, and the aims of this section are to: 
(i) highlight an unrealistic assumption in the literature underlying the current modelling of a single contact; 
(ii) introduce two more realistic models that improve predictability and optimisation of $\lt$; 
(iii) derive the explicit dependence of $\lt$ on the grain sizes and properties for all the models, making it possible to compare them. 

\begin{figure}[h]
  \includegraphics[width=0.45\textwidth]{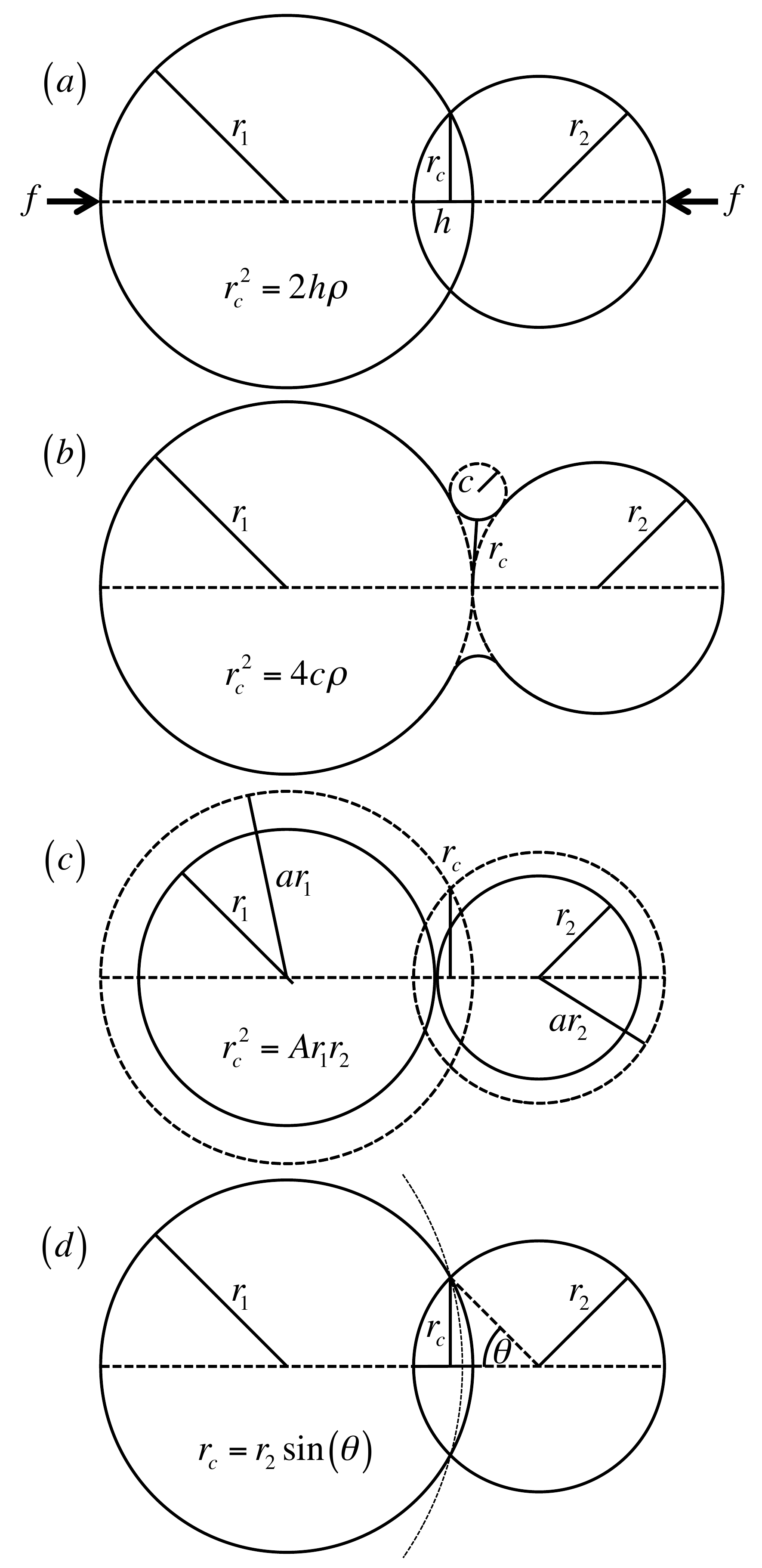}
  \caption{Four types of geometries to model intergranular contacts, and their respective radii of contact discs, $r_c$: (a) the H-model, based on a Hertzian contact force model; (b) the C-model, taking into consideration a curved interface due to sintering; (c) the geometric I-model, based on an inflation of grains due to sintering; and (d) the $\theta$-model, assuming a constant contact angle of the smaller grain. The thin line in (d) represents an even larger $r_1$, demonstrating that this model is unrealistic.}
  \label{fig:contact_types}
\end{figure}

The single contact TPB is the circumference of the contact disc, of radius $r_c$, between two grains of phases 1 and 2,
\begin{align} \label{eq:lt}
\lt = 2 \pi r_c \ .
\end{align}
Sintering theory \cite{Cob58} considers two geometric models for $r_c$ and a combination of them: a simple Hertzian overlap between the two grains, as in Fig. \ref{fig:contact_types}a, which we call H-model, and a curved transition layer between the grains, as in Fig. \ref{fig:contact_types}b, which we call C-model. The respective expressions for $r_c$ in terms of grain radii are (see details in appendix \ref{app:geometry}):
\begin{align}
r_{c, {\rm H}}^2 &= 2 h \rho + O(h^2) \ , \label{eq:r_overlap} \\
r_{c, {\rm C}}^2 &= 4 c \rho + O(c^{3/2}) \ , \label{eq:r_curvature}
\end{align}
where $h$ and $c$, shown in Fig. \ref{fig:contact_types}, are, respectively, the overlap between the grains and the curvature of the transition layer, and $\rho$ is the effective radius, $\rho = r_1 r_2 / (r_1 + r_2)$. When $r_1 \! = r_2 \! = r_0$, eqs. (\ref{eq:r_overlap}) and (\ref{eq:r_curvature}) reduce to $r_{c, {\rm H}}^2 \! = \! h r_0$ and $r_{c, {\rm C}}^2 \! = \! 2 c r_0$, in agreement with \cite{Cob58}.

In appendix \ref{app:omodel} we show that $h \propto \rho$ and, specifically,
\begin{align} \label{eq:rco}
r_{c, {\rm H}} = \frac{3 \sqrt{6} P_o} {E^*} \rho \equiv C \rho \ ,
\end{align}
where $P_o$ is the external pressure and $E^*$ is a function of the grains' elastic moduli and Poisson ratios under the sintering conditions. Both $P_o$ and $E^*$ are taken as constants in our analysis and thus the grain size dependence of $r_{c, {\rm H}}$ is only proportionally to $\rho$.

Since $\lt$ depends linearly on $r_c$ then $\Lt$ is proportional to $\rho$, as suggested in \cite{Pan98, Max03, Mar06}. 
In particular, Pan et al. \cite{Pan98} observed that the overlap between the two particles is not affected significantly by their size difference, as long as the difference is less than 50\%. This indicates that $h \propto \rho$, as we have found: when the grain sizes are comparable, $\rho$, and thus $h$, are almost constant. But when one grain is much smaller, its size dominates $\rho$, and thus $h$.

The other common model of sintering is a geometrical inflation algorithm, modelling grain growth, which we call I-model. It consists of multiplying all radii by a constant factor, $a$, as sketched in Fig. \ref{fig:contact_types}c \cite{Gol08, Ali08, Chen09}, neglecting intergranular contact forces. This model gives for the contact radius $r_{c, {\rm I}} = \sqrt{A r_1 r_2 - B (r_1^2 + r_2^2)}$, with $A = \frac{1}{2} (a^4 - 1)$ and $B = \frac{1}{4} (a^2 - 1)^2$ (see appendix \ref{app:geometry}). In practice $1 \le a \le 1.1$, translating to $B < A/20 \ll A$, and we can safely approximate
\begin{align} \label{eq:r_inflation_approx}
r_{c, {\rm I}} = \sqrt{A r_1 r_2} + O(B / A) \ ,
\end{align}
yielding that the I-model TPB length is proportional to the geometric mean, $\sqrt{r_1 r_2}$.

Another common estimate for $r_c$ in the literature is \cite{Co98, Chen09}
\begin{align} \label{eq:r_min}
r_{c,{\rm min}} = \min(r_1, r_2) \sin(\theta) \ ,
\end{align}
with $\theta$ the contact angle of the smaller grain, shown in Fig. \ref{fig:contact_types}d, the value of which is regarded as constant, typically $15^\circ$. This model, which we call $\theta$-model, is a crude approximation, but it appears reasonable because: (i) one does not expect $\theta$ to scale with $r$ and (ii) the smaller grain's contact angle is not too sensitive to the larger grain's size, at least when the size difference is large.

\begin{figure}[h]
  \includegraphics[width=0.5\textwidth]{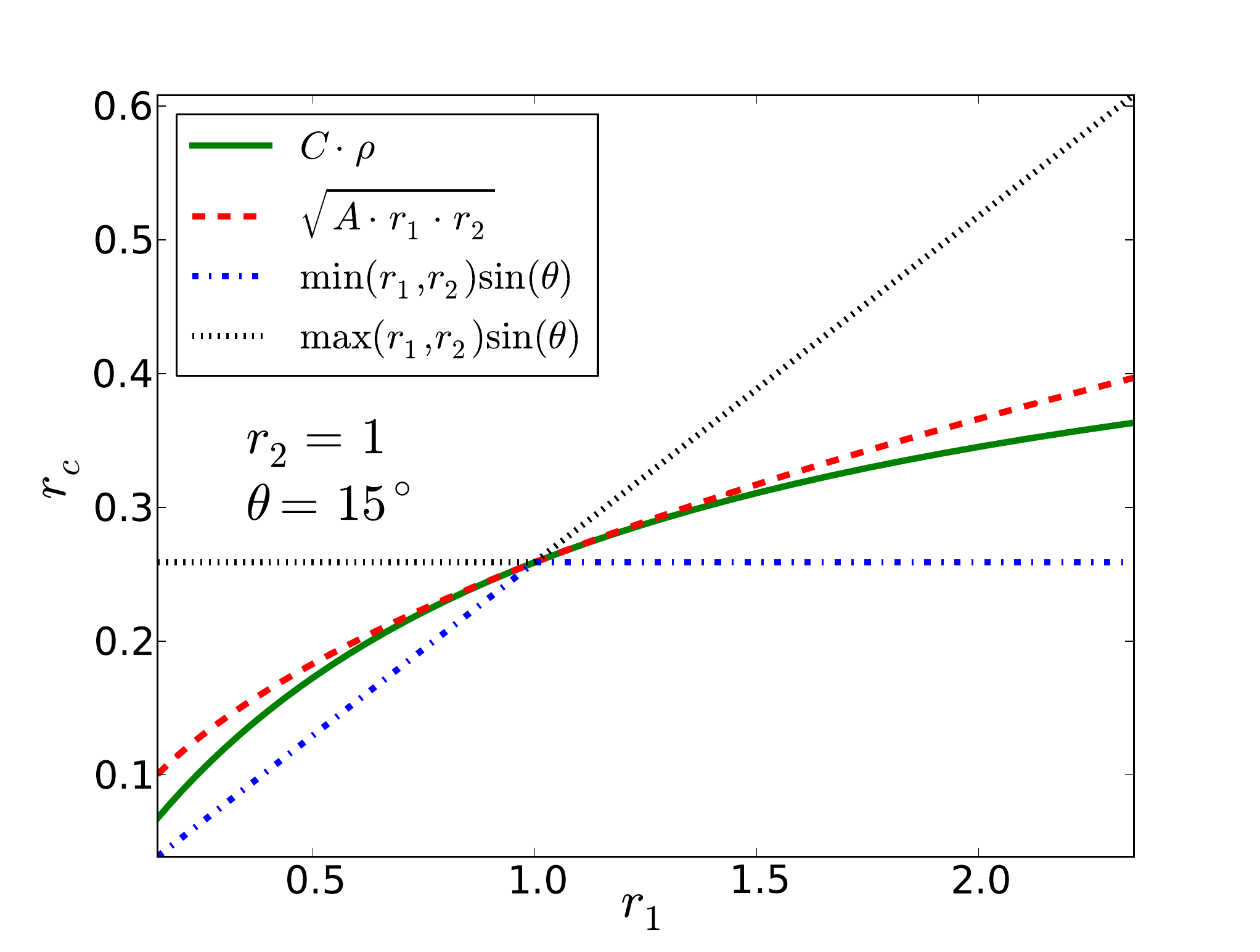}
  \caption{The values of the contact disc radius, $r_c$, given by the four contact models. The less realistic $r_{c,{\rm min}}$ and $r_{c,{\rm max}}$ bound, respectively, from below and from above, the values of the more realistic $r_{c, {\rm H}}$ and $r_{c, {\rm I}}$. Although $r_1 \ge r_2$ in our analysis, here we show also $r_1 < r_2$ for clarity.}
  \label{fig:four_single_tpbs}
\end{figure}

However, for studying the effect of particle size on TPB length, this approximation suffers from several disadvantages. Firstly, $r_{c,{\rm min}}$ is insensitive to the larger grain size, which is unphysical and introduces a large error, especially when $r_1 \approx r_2$. Secondly, the assumption of a constant $\theta$ leads to a reduction in the overlap between the grains upon an increase in the larger grain size, as can be observed in Fig. \ref{fig:contact_types}d. This is unphysical since the solid volume displaced by the grains pushing against one another must increase with both radii. In particular, the $\theta$-model does not satisfy the relation $h \propto \rho$.

Fig. \ref{fig:four_single_tpbs} summarises the above three models for the single TPB length, described by eqs. (\ref{eq:rco}), (\ref{eq:r_inflation_approx}) and (\ref{eq:r_min}), as well as the complementary of the latter, $r_{c,{\rm max}} = \max(r_1, r_2) \sin(\theta)$. The constants $A$ and $C$ in these relations were chosen to give the same value when $r_1 = r_2 = 1$. We note that the expressions $r_{c,{\rm min}}$ and $r_{c,{\rm max}}$ bound between them the values of $r_{c,{\rm H}}$ and $r_{c,{\rm I}}$, and we regard them as lower and upper bounds to the possible behaviours of $\lt$. Although $r_{c, {\rm H}}$ and $r_{c, {\rm I}}$ are more realistic and are derived more rigorously, without additional information about the mechanics of the sintering process, it is difficult to compare these models. Nevertheless, we emphasise that our results hold for all four models.

\begin{figure}[h]
  \includegraphics[width=0.5\textwidth]{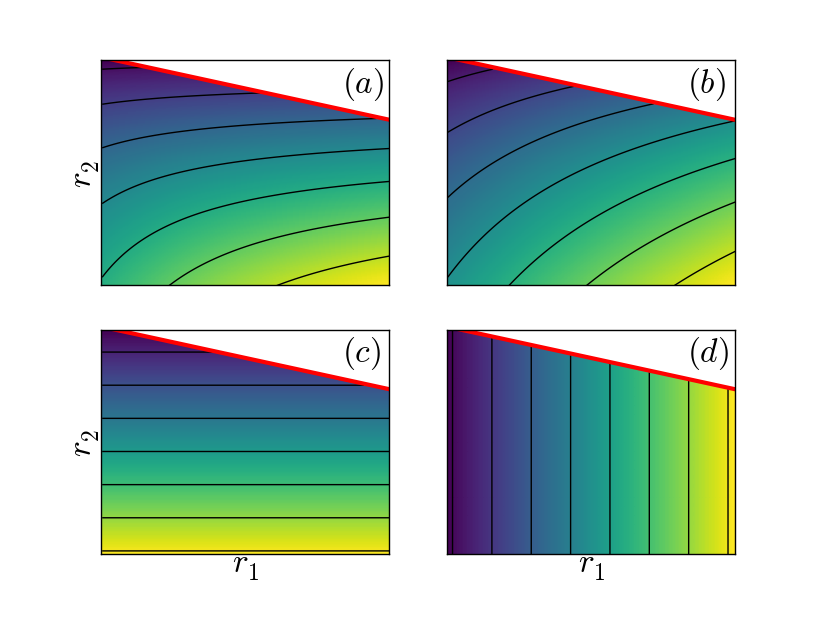}
  \caption{A single TPB length, $\lt$, as a function of the two grain radii, $r_1$ and $r_2$, (a) the H-model, $C = 0.518$; (b) the I-model, $a = 1.032$; (c) the $\theta$-model with the smaller radius, $\theta = 15^\circ$; (d) the $\theta$-model with the larger radius, $\theta = 15^\circ$. The axes, which have been removed for clarity, are the same as in Fig. \ref{fig:volume_fraction}. The value ranges of $\lt$ are: (a) [0.14-0.7], (b) [0.19-0.77], (c) [0.08-0.49] and (d) [0.49-1.22] $\mu$m. $\lt(r_0, r_0)$ (bottom left corner) is equal for all models ($0.488 \mu$m).}
  \label{fig:four_single_tpb_heat_maps}
\end{figure}

In Fig. \ref{fig:four_single_tpb_heat_maps} we show the dependence of $\lt$ on $r_1$ and $r_2$ for each of the four models. The H- and I-models are qualitatively similar, showing $\lt$ increasing monotonically with both radii. In the two $\theta$-models, $\lt$ only depends on one of the radii. Here, and in the following, we set $A = 0.067$, which corresponds to $a = 1.032$, and $C = 0.518$, which normalises all models to the same value of $\lt$ for $r_1 = r_2 = r_0$. We note that, for this value of $a$, $B < A / 60$, making the expansion to first order in eq. (\ref{eq:r_inflation_approx}) a very accurate approximation.

\section{The number of contacts} \label{sec:number_of_contacts}

The electrochemical reaction in an SOFC composite electrode occurs close to the interface between the two solid phases. Therefore, we are interested in the number of $1\hy 2$ contacts per unit volume. This number depends on all four parameters, $\{r_1, r_2, \psi_1, \phi \}$ and it can be written as
\begin{align} \label{eq:c_kl}
c_{1 \hy 2} = n_1 \cdot z_{1 \hy 2} \ ,
\end{align}
where $z_{i \hy j}$ is the mean number of contacts that one type-$i$ grain has with type-$j$ grains. In terms of the volume fractions,
\begin{align} \label{eq:n_1}
n_1 = \frac{\psi_1 (1 - \phi)}{\frac{4 \pi}{3} r_1^3} \ .
\end{align}
To obtain $z_{1 \hy 2}$, we use the estimate of Suzuki and Oshima \cite{SuOsh83}:
\begin{align} \label{eq:z_kl}
z_{1 \hy 2} = s_2 \cdot z_{1 \hy 2}^{(l)} \ ,
\end{align}
where $s_i$ is the fraction of the surface of type-$i$ grains of the entire solid surface area and $z_{1 \hy 2}^{(l)} = \lim_{\xi_2 \to 1} z_{1 \hy 2}$, namely $z_{1 \hy 2}$ in the limit of a very dilute concentration of type-1 grains, where, presumably, no contacts occur between type-1 grains. Like $\psi_i$, $s_i$ can be expressed in terms of the number fractions,
\begin{align}
s_i = \frac{\xi_i r_i^2}{\xi_1 r_1^2 + \xi_2 r_2^2} \ .
\end{align}
Suzuki and Oshima then approximate the value of $z_{1 \hy 2}^{(l)}$ as
\begin{align} \label{eq:z12limit}
z_{1 \hy 2}^{(l)} = \frac{(2 - \sqrt{3}) (P + 1) N_c}{2 [1 + P - \sqrt{P (P + 2)}]} \ ,
\end{align}
where $N_c$ is the mean number of contacts per grain in a randomly packed monodisperse system. In \cite{Suz81, Ziff17}, $N_c$ is assumed to be 6 and we adopt this assumption. However, the choice of $N_c$ only affects our calculation of the TPB length by a constant and, therefore, it leaves unchanged the optimal set of parameters that we identify.

Eqs. (\ref{eq:number_to_volume_fraction}) and (\ref{eq:c_kl})-(\ref{eq:z12limit}) enable us to express $c_{1 \hy 2}$ in terms of $r_1$, $r_2$, $\psi_1$ and $\phi$. In this solution, $z_{1 \hy 2}^{(l)}$, the average number of type-2 grains surrounding a lone type-1 grain, depends only on $P$ and $N_c$. As expected, it increases monotonically from $N_c$, when $P=1$, to about $15 N_c$ when $P \to P_{\rm max}$. $z_{1 \hy 2}$ depends on the population of type-1 grains as well, dropping from $z_{1 \hy 2}^{(l)}$ as $\xi_1$ increases. This is because of the increased occurrences of $c_{1\hy 1}$ contacts, which prohibit $c_{1\hy 2}$ contacts.

\begin{figure}[h]
  \includegraphics[width=0.47\textwidth]{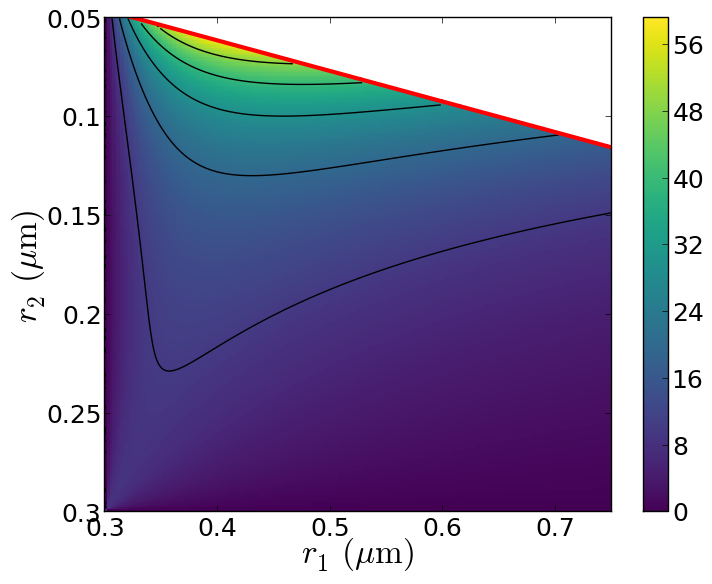}
  \caption{The number of 1-2 contacts per $1 \mu {\rm m}^3$, $c_{1 \hy 2}$, as a function of the two grain radii, $r_1$ and $r_2$. The values of $\psi_1$ are obtained using the same fixed value of $R_h$ as in Fig. \ref{fig:volume_fraction}.}
  \label{fig:c_kl}
\end{figure}

In Fig. \ref{fig:c_kl} we plot the solution for $c_{1 \hy 2}$ as a function of $r_1$ and $r_2$ for a fixed value of $R_h$. It reflects the balance between increasing $n_1$, which requires increasing $\psi_1$ by reducing $r_1$ (see eq. (\ref{eq:n_1}) and Fig. \ref{fig:volume_fraction}), and increasing $z_{1 \hy 2}$, which requires increasing $P$ (see eq. (\ref{eq:z12limit})). We find that there is an overall maximal number of 1-2 contacts around $r_1 = 0.4 \mu$m. This maximum is at the smallest $r_2$ possible, as both $n_1$ and $z_{1 \hy 2}$ increase with decreasing $r_2$.

\section{The active TPB} \label{sec:active_fraction}

The function of the electrodes is to conduct electrons and ions to and from the TPB, where the electrochemical reaction takes place. To this end, single grains of each phase must form a cluster that percolates between the boundaries of the electrode \cite{Co98}. The 1-2 contacts connecting such grains are called active and their number can be estimated as
\begin{align} \label{eq:active_c_kl}
c_{1 \hy 2}^{\rm (a)} = p_1 \cdot p_2 \cdot c_{1 \hy 2} \ ,
\end{align}
where $p_i$ is the probability that a type-$i$ grain is a part of the percolating cluster of type-$i$ grains, and the two probabilities are assumed to be independent. $p_i$ is primarily determined by the coordination number within the phase, $z_{i \hy i}$. We use an existing approximation in the literature \cite{Bou91, Chen09, Zhu08, Ber11Comp},
\begin{align} \label{eq:approxpi}
p_i = \left[ 1 - \left( \frac{\alpha - z_{i \hy i}}{\beta} \right)^\gamma \right]^\delta \ ,
\end{align}
for which different studies fit different sets of parameters. We choose to use the parameters suggested in the comparative study \cite{Ber11Comp}: $\alpha = 4.236$, $\beta = 2.472$, $\gamma = 3.7$, and $\delta = 1$, both because they give the correct value of $z_{i,i}$ at the percolation threshold, $z_{i \hy i}^{\rm (th)} = 1.764$ \cite{Kuo95}, and they describe well the simulated data of \cite{Bou91}. We reiterate that the particular choice of parameters and the exact dependence of $p_i$ on $z_{i \hy i}$ do not affect our final conclusions.

Similarly to eq. (\ref{eq:z_kl}), we hypothesize that $z_{i \hy i} = s_i \cdot z_{i \hy i}^{(l)}$, but since $z_{i \hy i}^{(l)} = \lim_{\xi_i \to 1} z_{i \hy i} = N_c$, we have
\begin{align}
z_{i \hy i} = s_i \cdot N_c \ ,
\end{align}
which can be substituted in (\ref{eq:approxpi}) to give $p_i$.

\begin{figure}[h]
  \includegraphics[width=0.5\textwidth]{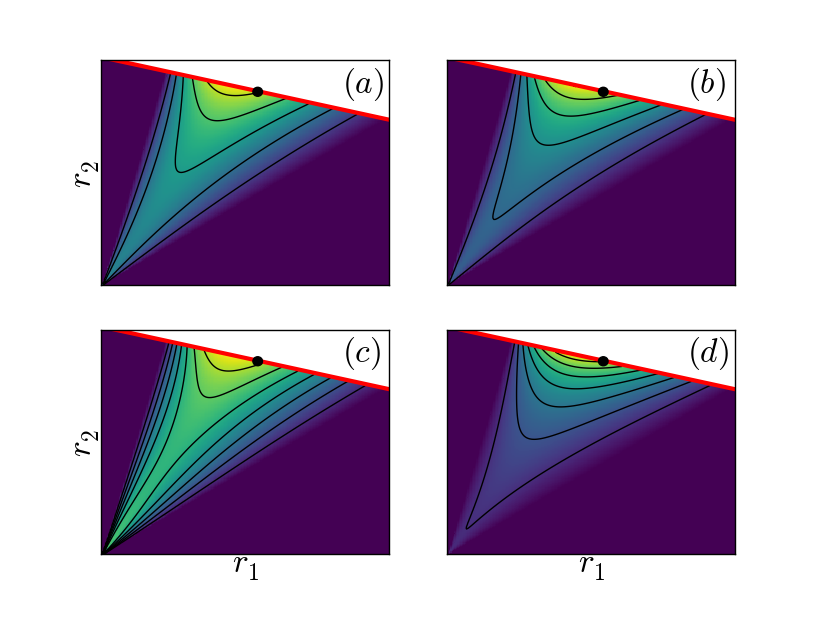}
  \caption{The total active TPB length per $\mu {\rm m}^3$, $\Lt$, as a function of the two grain radii, $r_1$ and $r_2$ for the four contact models and for the same value of $R_h = 54$nm. The single contact TPB length is calculated for: (a) the H-model, $C = 0.518$; (b) the I-model, $a = 1.032$; (c) the $\theta$-model with the smaller radius, $\theta = 15^\circ$; (d) the $\theta$-model with the larger radius, $\theta = 15^\circ$. The axes, which have been removed for clarity, are the same as in Figs. \ref{fig:volume_fraction} and \ref{fig:c_kl}. The highest values of $\Lt$ are obtained at $(r_1, r_2) = (0.500 \mu {\rm m}, 0.077 \mu {\rm m})$ and their values are: (a) 8.95, (b) 13.14, (c) 5.17 and (d) 33.40 $\mu {\rm m}^{-2}$. The value of $\Lt(r_0, r_0)$ (bottom-left corner) is equal for all contact models ($\simeq 3.53 \mu {\rm m}^{-2}$). The black dots are at $(r_1^{\rm (max)}, r_2^{\rm (max)}) = (0.543 \mu {\rm m}, 0.084 \mu {\rm m})$, which are the best grain size pair that can be compared fairly to $(r_0, r_0)$, and they correspond to $\Lt$ = (a) 7.60, (b) 11.15, (c) 4.39 and (d) 28.36 $\mu {\rm m}^{-2}$.}
  \label{fig:total_active_tpb}
\end{figure}

We can now estimate the total active TPB density:
\begin{align}
\Lt = \lt \cdot c_{1 \hy 2}^{\rm (a)} \ ,
\end{align}
where $c_{1 \hy 2}^{\rm (a)}$ is obtained from eq. (\ref{eq:active_c_kl}) and $\lt$ is obtained from eq. (\ref{eq:lt}) for any one of the four models discussed in section \ref{sec:single_tpb}. In Fig. \ref{fig:total_active_tpb} we plot $\Lt$ as a function of $r_1$ and $r_2$ for each contact model. All four plots feature large regions of $\Lt=0$ that arise because one of the phases does not percolate between opposite boundaries. A key observation is that, within the `percolating band', $\Lt$ increases monotonically with increasing $r_1$ and with decreasing $r_2$, for all four models. Significantly, $\Lt$ reaches a maximum when $P = P_{\rm max}$ for all the single contact models. The pair of radii giving the longest TPB is $(r_1, r_2) = (0.500, 0.077)\mu$m. However, we shall focus on the pair $(r_1^{\rm (max)}, r_2^{\rm (max)}) = (0.543, 0.084)\mu$m, marked by a black dot in Figs. \ref{fig:total_active_tpb}a-d, for a reason that will become clear below. It is important to emphasise that, keeping all other variables constant and changing only $r_0$, these values would simply scale linearly:
\begin{align} \label{eq:size_scale}
\left(r_1^{\rm (max)},r_2^{\rm (max)}\right)=\left(1.81,0.28\right) r_0 \ .
\end{align}
For the H- and I-models (Figs. \ref{fig:total_active_tpb}a and \ref{fig:total_active_tpb}b) the maximum values of $\Lt$ are about 2.5 and 3.7 times as high as $\Lt(r_0, r_0)$, respectively.

\begin{figure}[h]
  \includegraphics[width=0.5\textwidth]{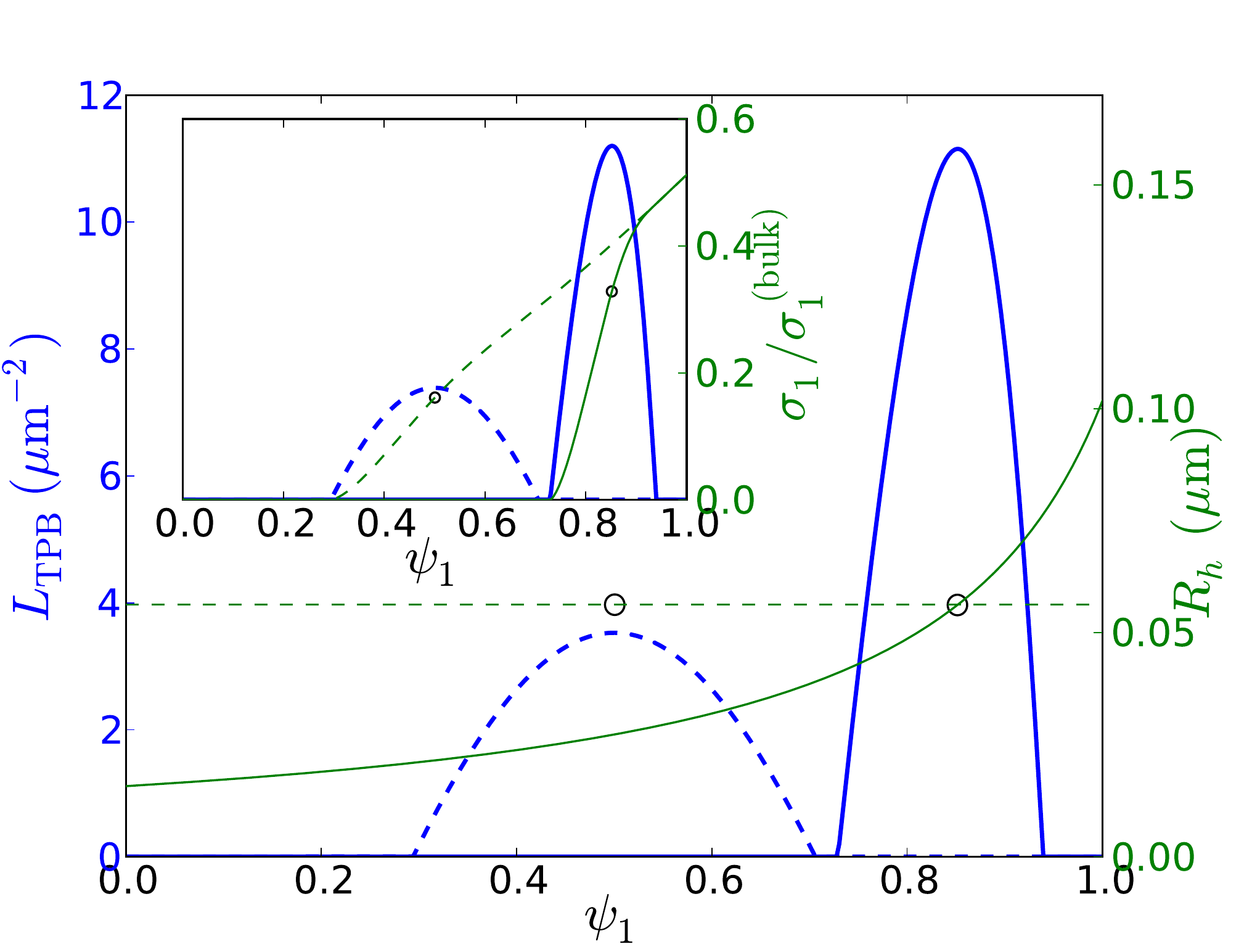}
  \caption{Comparison of a monodisperse system with grain radius $r_0 = 0.3 \mu$m (dashed lines) and a bidisperse system with grain radii $0.543 \mu$m and $0.084 \mu$m (solid lines), for the I-model. The value of $\Lt$ is in thick blue lines and $R_h$ in thin green lines. The maximum value of $\Lt$ for the bidisperse mixture is 3.16 times as high. At the peak, $R_h$ is the same for both mixtures (black circles). Inset: The effective conductivity of type-1 grains for the two mixtures (thin green lines). At the peak, the effective conductivity of the bidisperse mixture is twice as high (black circles).}
  \label{fig:vary_psi}
\end{figure}

To highlight the effect of the size distribution, we compare in Fig. \ref{fig:vary_psi} the dependencies of $\Lt$ and $R_h$ on $\psi_1$ for the pairs $(r_1^{\rm (max)}, r_2^{\rm (max)})$ and $(r_0, r_0) = (0.3 \mu{\rm m}, 0.3 \mu{\rm m})$ within the I-model. The highest value of $\Lt$ is 3.16 times as high for the bidisperse system, obtained at $\psi_1^{\rm (max)} = 85\%$. While $R_h$ is constant for the pair $(r_0, r_0)$, it changes monotonically with $\psi_1$ for $(r_1^{\rm (max)}, r_2^{\rm (max)})$ -- the higher the concentration of 1-type grains, the larger $R_h$. We draw attention to the fact that $R_h$ is the same at the respective maxima; this would not have been the case had we used the maximum TPB pair from Fig. \ref{fig:total_active_tpb}, $(0.500 \mu {\rm m}, 0.077 \mu {\rm m})$. This is because fixing $(r_1, r_2)$ and varying $\psi_1$ moves the system away from the surface shown in Fig. \ref{fig:total_active_tpb}, thus changing $R_h$. For example, varying $\psi_1$ at $(r_1, r_2) = (0.500 \mu {\rm m}, 0.077 \mu {\rm m})$, $\Lt$ could be increased even further, but this would be at the expense of reducing $R_h$, resulting in an unfair comparison. The volume fraction $\psi_1$ that gives the highest value of $\Lt$ for the pair $(r_1^{\rm (max)}, r_2^{\rm (max)})$ is the same one as in Fig. \ref{fig:total_active_tpb}.

We carried out the same analysis for the two $\theta$- and H-models. For all contact models, $\Lt$ is maximised at the same values of $\{ r_1^{\rm (max)}, r_2^{\rm (max)},\psi_1^{\rm (max)} \}$, but its value at the maximum changes. For all models, we obtain significant improvements by: $\Lt^{\rm (max)} / \Lt^{(0)} = 1.24, 8$ and $2.15$, respectively. We conclude that using grains of different sizes, with the appropriate volume fraction, increases the length of the active TPB regardless of the specific single contact model. Nevertheless, different models lead to different improvements. Since the H- and I-models are the more realistic ones, we would expect an actual improvement by a factor of 2.15--3.16.

\section{Consequences for other features} \label{sec:other_features}

As mentioned, the performance of SOFCs depends on a number of properties and it is important to know how the maximisation of the TPB length affects them. One essential such property is the ionic conductivity. We compare the effective ionic conductivities of the monodisperse and the TPB-maximising bidisperse mixtures. To this end, we use the Bruggeman's model-based effective medium result \cite{Chen09}
\begin{align} \label{eq:eff_cond}
\frac{\sigma_i}{\sigma_i^{\rm (bulk)}} = \big[ (1 - \phi) \cdot \psi_i \cdot p_i \big]^\mu \ ,
\end{align}
where $\sigma_i$ is the conductivity of the type-$i$ phase and the power was fitted from measurements at $\mu=1.5$. Note that the term in brackets is the volume fraction of the percolating part of phase $i$, out of the whole domain.

The ion-conducting phase is typically the least-conducting and, to achieve best performance, it should consist of the larger grains, thus we assign it to type-1. With $\psi_1^{\rm (max)} > 0.5$, this guarantees higher effective conductivity for bidisperse systems than the alternative. Indeed, Figs. \ref{fig:intro} and \ref{fig:vary_psi} (inset) show that the effective conductivity of the type-1 phase is twice as high for $\{ r_1^{\rm (max)}, r_2^{\rm (max)}, \psi_1^{\rm (max)} \}$, compared to $\{ r_0, r_0, 1/2 \}$. This is irrespective of the single contact model. Thus, we conclude that maximising $\Lt$ by using the appropriate bidisperse powder, of which the larger grains are ion-conducting, also increases the effective ionic conductivity without compromising the mean pore size. The trend of the microstructure we offer here, of large ion-conducting grains surrounded by smaller electron-conducting ones, is consistent with the recently proposed method of impregnated nano-structured electrodes \cite{Jia12}, although the latter requires sophisticated fabrication techniques that are difficult to scale up.

Fig. \ref{fig:vary_psi} shows that the range of values of $\psi_1$, for which $\Lt$ is non-zero in the optimal bidisperse system, is about half that for the monodisperse system. Although this requires more accuracy in the electrode production process, which could be a potential drawback, the accuracy of the volume fraction under the now-standard advanced techniques of powder weighing, mixing and preparation, is better than 1\%, even when allowing for spatial variations. Therefore, our predicted improvement to the TPB density is well within the current technological capabilities.

We also need to consider the consequences of the optimisation on spatial variations of the hydraulic radius. In our optimal solution, for which the mean hydraulic radius is $R_h = 54$nm, the small grains are more numerous ($\xi_2 = 98\%$) and we need to check that there are no large regions, where the hydraulic radius is at its minimum value, $R_h^{\rm (min)} = \frac{\phi r_2}{3 (1 - \phi)} = 16{\rm nm}\approx 30\% R_h$, which might restrict considerably gas flow. However, this is not a problem because:
(i) most of the volume ($\psi_1 = 85\%$) is occupied by the large grains and, therefore, $R_h$ holds locally within the lion's share of the volume;
(ii) the less accessible regions are also the less important ones for the operation of the SOFC because they contain no type-1 grains, prohibiting the electrochemical reaction anyway.
To see this, we consider the mean distance between type-1 grains and define as $r_1^{\rm (full)}$ the effective radius that these would have, had they occupied the entire solid volume:
\begin{align}
\frac{4 \pi}{3} n_1 (r_1^{\rm (full)})^3 = (1 - \phi) \ .
\end{align}
For the random packing porosity $\phi = 0.36$, $r_1^{\rm (full)} = 1.05 r_1^{\rm (max)}$, which means that the mean distance between the surfaces of nearest-neighbour type-1 grains is $d_1 = 2(r_1^{\rm (full)} - r_1^{\rm (max)}) = 0.1 r_1^{\rm (max)}$. Namely, one cannot fit even one type-2 grain between them, indicating that the aforementioned regions, with no type-1 grains, are limited. Furthermore, the number of contacts between type-1 grains at $\psi_1^{\rm (max)}$ is $z_{1 \hy 1} = 2.8$, also pointing to good proximity.

\section{A guide for the electrode design} \label{sec:recipe}

We can now assemble our results into a guide for the design of SOFC electrodes. \\
1. Start with the desired, or given, values of $\phi$ and $R_h$. $\phi$ may be dictated by the packing and sintering protocols and $R_h$ by a desired limitation on the maximal Knudsen number and the mean free path of the reactants/products. \\
2. Identify the monodisperse system that the values of $\phi$ and $R_h$ correspond to, $r_0 = 3 (1 - \phi) R_h / \phi$. \\
3. Use relation (\ref{eq:size_scale}), which holds for any values of $\phi$ and $R_h$, to determine the grain sizes and $\psi_1 = 85\%$.

\begin{figure}[h]
  \includegraphics[width=0.5\textwidth]{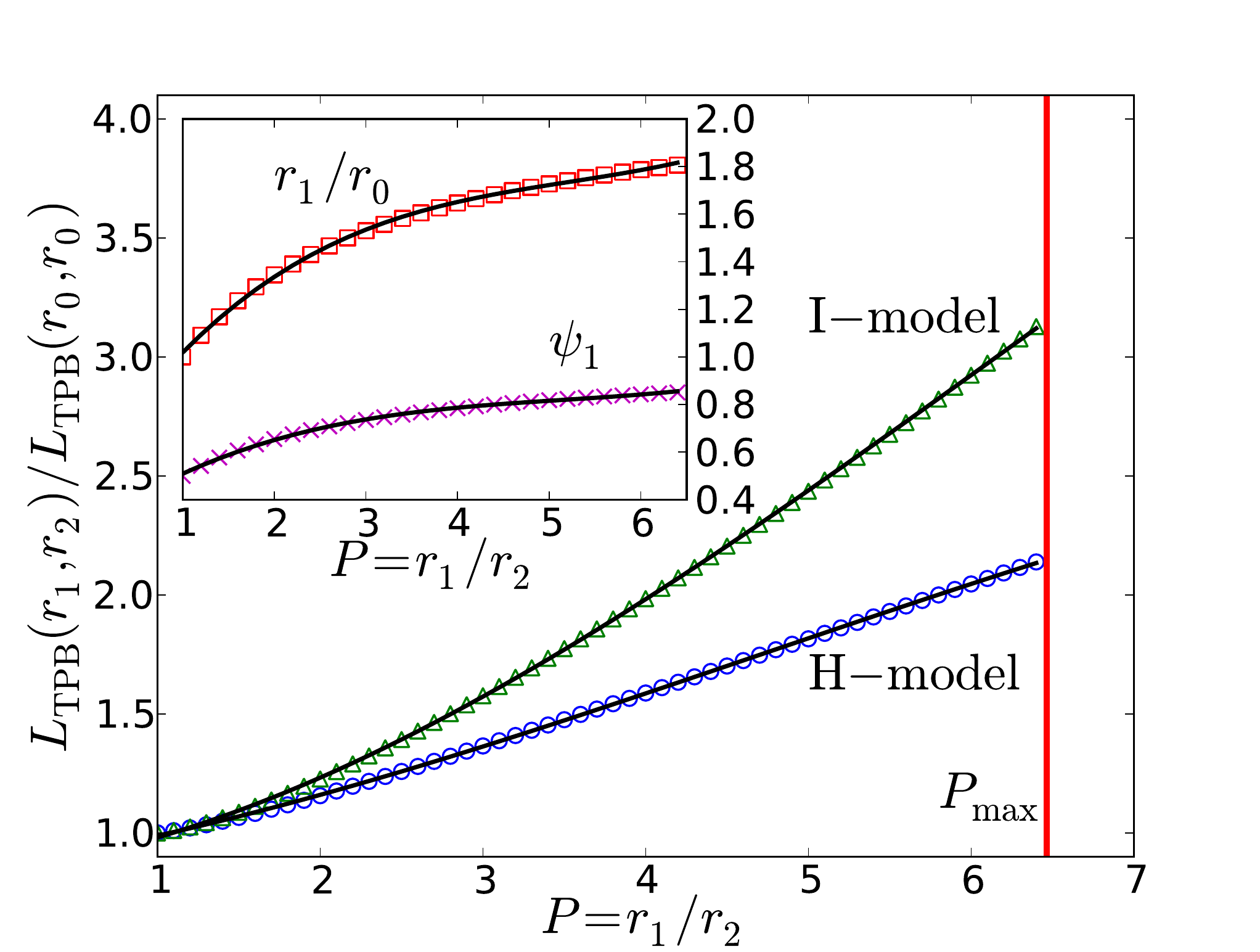}
  \caption{The relative improvement, $\Lt^{\rm (max)} / \Lt^{(0)}$, vs. grain size ratio, $P = r_1 / r_2$ for the H-model (blue circles) and I-model (green triangles). The vertical thick red line marks the maximum possible ratio, $P_{\rm max}$. 
Inset: The values of $r_1/r_0$ (red squares) and $\psi_1$ (magenta crosses), for which $\Lt$ is maximal, vs. $P$. The black lines are cubic polynomial fits.}
  \label{fig:recipe}
\end{figure}

However, other considerations may dictate a size ratio lower than $P_{\rm max}\simeq 6.46$, e.g. for better mixing. To facilitate a general guide we repeated the above analysis for any value of $1\leq P \leq P_{\rm max}$. Namely, for each value of $P$ we determined the values of $r_1$ and $r_2$ that maximise $\Lt$ and calculated the improvement in its value. The results are summarised in Fig. \ref{fig:recipe}, which shows, for any $P$, the value of $r_1$, $\psi_1$ and $\Lt^{\rm (max)} / \Lt^{(0)}$ for both the H- and I-models. To assist with the choice of parameters, we provide the following fits, also shown in Fig. \ref{fig:recipe}:
\begin{align}
r_1^{\rm (max)} / r_0 &\simeq 0.0062 P^3 - 0.097 P^2 + 0.565 P + 0.54 \ , \nonumber \\
\psi_1^{\rm (max)} &\simeq 0.0029 P^3 - 0.045 P^2 + 0.256 P + 0.30 \ . \nonumber
\end{align}
These, together with $r_2 = r_1 / P$, give the ideal grain sizes and their relative concentrations for any required set of values $R_h, \phi$ and $P$. Fig. \ref{fig:recipe} makes it evident that the larger the value of $P$, the longer the overall TPB, in contrast to most current guidelines in the literature for the design for SOFC electrodes.

\section{Numerical support} \label{sec:numerical}

To support our results, we generated seven composite electrodes, using the sedimentation algorithm presented in \cite{Ber12,SuOsh83}, to validate the analytical model (\ref{eq:z_kl})-(\ref{eq:z12limit}) for the number of contacts. The algorithm simulates rigid spherical particles dropped sequentially from random positions at the top of a prismatic domain. A particle comes to rest either on the floor or when it is in contact with three stationary particles. The particle's position is fixed and a new particle is deposited. The desired volume fraction, equal to the optimal $\psi_1$ for $1\leq P\leq 4$, is enforced by assigning a weighted probability to the particle selection. Once the domain is entirely filled, particles are inflated according to the I-model and the resulting active TPB density is calculated. 
The results are summarised in Fig. \ref{fig:intro} and in Table \ref{tab:compare1}: all measured values of $\Lt$ surpass our predictions by 10\%--16.5\%, and all improvement ratios agree with our predictions to an accuracy of 6\%. For example, the simulated system with $P = 4$ and $\psi_1 = 78\%$ has TPB density 2.02 as high as $P = 1$ -- within 2\% of our predictions.

\begin{table}[h]
\centering
\begin{tabular}{ | c | c | c | c | c | }
\hline
$\Lt$ ($\mu$m$^{-2}$) & $P = 1$ & $P = 2$ & $P = 3$ & $P = 4$ \\
\hline
analitical & 3.5283 & 4.3267 & 5.5623 & 7.0024 \\
(imp. factor) & 1.00 & 1.23 & 1.58 & 1.98 \\
\hline
numerical & 3.8779 & 4.9865 & 6.4803 & 7.6622 \\
(imp. factor) & 1.00 & 1.29 & 1.67 & 1.98 \\
\hline
\end{tabular}
\caption{Comparison of the analytical and numerical results for $\Lt$, in four systems of grain size ratios $1\leq P\leq 4$. The lines ``imp. factor'' report the TPB density ratio between the respective system and the one of $P = 1$.}
\label{tab:compare1}
\end{table}

We also calculated the effective conductivities according to \cite{Ber13}. The results are summarised in Table \ref{tab:compare2}. The numerical values are lower than our estimates because relation (\ref{eq:eff_cond}) overestimates the conductivity of granular packs, as was also observed in \cite{Chu14, San10}. Specifically, the small size of the contact discs is not taken into consideration in relation (\ref{eq:eff_cond}). Therefore, that relation provides an upper bound for the conductivity. Nevertheless, the improvement ratios are larger than our analysis predicts. This is because the benefit in having large grains, in terms of conductivity, is greater than their mere volume, since they are also harder to displace by the small electron-conducting grains, thus making the percolating path between opposite sides of the system less tortuous.

\begin{table}[h]
\centering
\begin{tabular}{ | c | c | c | c | c | }
\hline
$\sigma_1 / \sigma_1^{\rm (bulk)}$ & $P = 1$ & $P = 2$ & $P = 3$ & $P = 4$ \\
\hline
analitical & 0.161 & 0.233 & 0.272 & 0.296 \\
(imp. factor) & 1.00 & 1.45 & 1.69 & 1.85 \\
\hline
numerical & 0.0367 & 0.0624 & 0.107 & 0.134 \\
(imp. factor) & 1.00 & 1.70 & 2.91 & 3.66 \\
\hline
\end{tabular}
\caption{Comparison of the analytical and numerical results for the effective ionic conductivity, $\sigma_1 / \sigma_1^{\rm (bulk)}$, in four systems of grain size ratios $1\leq P\leq 4$. The lines ``imp. factor'' report the conductivity ratio between the respective system and the one of $P = 1$.}
\label{tab:compare2}
\end{table}

Thus, our numerical simulations confirm that a bidisperse mixture of the right volume fractions can increase significantly both the active TPB density and the effective ionic conductivity, supporting the main theoretical results of this study.
The excellent agreement between the numerical and analytical results establishes the intuitive relation (\ref{eq:z_kl}), proposed by Suzuki and Oshima \cite{SuOsh83}, combined with (\ref{eq:z12limit}), as a good effective-medium estimate of the number of contacts between particles of different sizes.

\section{Conclusions} \label{sec:conclusions}

We calculated the length of the TPB per unit volume, $\Lt$, in composite SOFC electrodes as a function of the volume fractions and the sizes of the grains, of which it is sintered, for a given mean hydraulic radius, $R_h$, with the aim to find the parameters that maximise it. We found that, in general, the TPB is always larger for bidisperse mixtures of large ion-conducting grains and small electron-conducting grains than for a monodisperse mixture of the same $R_h$. Our main result is that the longest TPB is obtained for a radius ratio of $P \simeq 6.46$ and that it is 2.15--3.16 times as large as the equivalent monodisperse mixture. Furthermore, we found that maximising $\Lt$ also doubles the ionic conductivity, improving performance further. These results indicate that the practices used currently in this field can be improved significantly. The TPB density was found to be sensitive to the relative volume fractions of the two phases, but this sensitivity can be readily handled by the standard composition preparation techniques. Although the ideal system consists of many more small grains than large ones, the larger volume occupied by the latter ensures that the mean hydraulic radius is sufficient for effective gas transport through the electrode to the reactive sites.

We also critiqued the commonly used approximation for a single contact TPB length, $\lt = 2 \pi \min (r_1, r_2) \sin \theta$, and showed that it is unrealistic and inadequate for analysing grain size effects. We proposed two other, more physically sound, models. Nevertheless, we showed that our overall conclusions on the best parameters to use are the same regardless of the single contact model.

We then extended our analysis to apply for any value of $R_h$ and $P$ and presented a clear method to identify the best choice of parameters, given a lower bound on the former and an upper bound on the latter. We also provided an explicit formula for the best choice of parameters in terms of the required size ratio.

Finally, we carried out numerical simulations, which support our analysis. In particular, the TPB density in a computer-generated composite electrode with $P = 4$ and $\psi_1 \simeq 78\%$ was found to be 2.02 as high as a monodisperse one of the same $R_h$ -- within 2\% of our predicted improvement. The improvement factor in effective conductivity was found to be up to 98\% as high as our prediction, which we explained. We look forward to further numerical and experimental tests of this analysis.

\section*{Acknowledgements}

SA acknowledges financial support by the Alan Howard Scholarship. AB acknowledges funding from the European Union's Horizon 2020 research and innovation programme under the Marie Sk{\l}odowska-Curie grant agreement No 654915.

\patchcmd{\appendices}{\quad}{. \ \ }{}{}
\numberwithin{equation}{section}
\patchcmd{\theequation}{.}{}{}{}

\begin{appendices}

\section{Calculating the radius of a single contact disc} \label{app:geometry}

\noindent 1. {\it H-model} \\

Defining $r_{c,\alpha}$ as the radius of the contact disc between two grains of radii $r_i$ and $r_j$ in the $\alpha$-model, where $\alpha =$ H, I or C, and assuming that the two grains remain spherical on contact \cite{Weis07}, we have
\begin{align} \label{eq:sphere_overlap}
r_{c, {\rm H}}^2 = \frac{4 d^2 r_i^2 - (d^2 - r_j^2 + r_i^2)^2}{4 d^2} \ ,
\end{align}
with $d = r_i + r_j - h$ the distance between the spheres centres. Substituting for $d$ and expanding in powers of $h$ we obtain eq. (\ref{eq:r_overlap}):
\begin{align}
r_{c, {\rm H}}^2 = 2 \rho \cdot h - \left( 1 - \frac{3 \rho}{r_i + r_j} \right) \cdot h^2 + O(h^3) \ .
\end{align}
While high-order terms do not necessarily depend on the radii through the combination $\rho$, they are negligible for $h\ll {\rm min}\{r_i,r_j\}$. For example, for the commonly used angle (see text), $\theta = 15^\circ$, corresponding to $h \simeq r_2 / 15$, the first order term is $99.14\%$ accurate for $r_i=r_j$ and $98.7\%$ accurate for $r_i= P_{\rm max}r_j$. \\

\noindent 2. {\it C-model} \\

Let us label $K$, $L$, $C$ and $O$, respectively, the centres of: grain $i$, grain $j$, the `sintering curvature', and the contact point (see Fig. \ref{fig:contact_types}b). Consider the triangle $LKC$, the lengths of the sides of which are: $R_i = r_i + c$, $R_j = r_j + c$ and $R = r_i + r_j$, and the line $OC$, which is of length $X = r_{c, {\rm C}} + c$. Using the cosine theorem for triangles $LKC$ and $OKC$, we obtain
\begin{align}
R_j^2 &= R_i^2 + R^2 - 2 R_i R \cos(\theta_i) \ , \\
X^2 &= R_i^2 + r_i^2 - 2 R_i r_i \cos(\theta_i) \ ,
\end{align}
with $\theta_i = \sphericalangle L \hy K \hy C$. Eliminating $\cos(\theta_i)$, expressing all, first in terms of $r_i$ and $r_j$ and then in terms of $\rho \equiv r_ir_j/(r_i+r_j)$, we get
\begin{align}
X^2 = 4 c \rho + c^2 \ ,
\end{align}
or
\begin{align}
r_{c, {\rm C}} = \sqrt{4 c \rho + c^2} - c \ .
\end{align}
Thus, in this model, $r_{c, {\rm C}}$ depends on the grain sizes only through $\rho$. Expanding in powers of $c / \rho$ gives eq. (\ref{eq:r_curvature}):
\begin{align}
r_{c, {\rm C}}^2 = 4 \rho \cdot c - 4 \sqrt{\rho} \cdot c^{3/2} + 2 \cdot c^2 + O(c^{5/2}) \ .
\end{align} \\

\noindent 3. {\it I-model} \\

Scaling the grain radii by $a>1$, eq. (\ref{eq:sphere_overlap}) applies with $r_i, r_j \to a r_i,ar_j$ and $d = r_i + r_j$, which can be manipulated into the form
\begin{align}
r_{c, {\rm I}}^2 = \left[ \frac{a^4 - 1}{2} \right] \cdot r_i \cdot r_j - \left[ \frac{(a^2 - 1)^2}{4} \right] \cdot (r_i^2 + r_j^2) \ .
\end{align}
Taking the square root of both sides yields the result in the main text. For $a = 1.032$, the first order term for $r_{c, {\rm I}}$ is 98.4\% accurate for $r_i = r_j$ and 94.4\% for $r_i = P_{\rm max} r_j$.

\section{The radius of a single contact disc is a function of grain radii only (H-model)} \label{app:omodel}

In this mean field approximation, we assume that the particles are uniformly distributed in the volume of the system. Let the mean coordination number per grain be $\bar{z}$ and the number of type-$i$ grains ($i = 1,2$) be $N_i = \xi_i N$, such that $\xi_1 + \xi_2 = 1$. The overall number of contacts is thus $\bar{z} N / 2$. The system is presumed cubic of size $L \times L \times L$ and under external pressure $P_o$, such that on each boundary we have a force $F = P_o L^2$.

Consider a thin sheet passing through the bulk system in parallel to, say, the $x$-boundary. The sheet is constructed such that it contains in its volume one layer of contacts. This `contact sheet' need not be exactly planar, but it may weave slightly to capture contacts that are within one grain radius away from a flat plane. On average, this layer contains
\begin{align}
z_w = \left(\frac{\bar{z} N}{2} \right)^{2/3}
\end{align}
contacts. The probability, $p_{ij}$, of contacts between type-$i$ and type-$j$ grains within such a uniform distribution is
\begin{align}
p_{11} = \xi_1^2 \ , \ \ p_{22} = \xi_2^2 \ , \ \ p_{12} = 2 \xi_1 \xi_2 \ ,
\end{align}
and, therefore, the mean number of contacts between type-$i$ and type-$j$ grains within the contact sheet is
\begin{align}
z_{w,ij} = p_{ij} z_w = p_{ij} \left( \bar{z} N / 2 \right)^{2/3} \ .
\end{align}
The total area of these contacts, projected in the $x$-direction, is
\begin{align}
s_{x,ij} = {z}_{w,ij} \epsilon \pi r_{c, {\rm H}}^2 \ ,
\end{align}
where $\epsilon = \frac{1}{\pi}\int_{-\pi/2}^{\pi/2} \cos{\theta} d\theta = 2/\pi$ arises from averaging over the distribution of the orientations of the $i \hy j$ contact discs. In the mean field approximation, we assume that the $x$-component of the force on the contact sheet area is distributed uniformly across the $y-z$ plane and, therefore, the force on all the $i \hy j$ contacts is $F_{x,ij}=s_{x,ij}F/L^2$. Dividing this force equally over all the $i \hy j$ contacts, the mean $x$-component of the force per contact is
\begin{align} \label{eq:fij}
f_{x,ij} = \frac{F_{x,ij}} {{z}_{w,ij}} = \epsilon \pi r_{c, {\rm H}}^2 P_o \ .
\end{align}
Assuming isotropic pressure on the system and carrying out the same analysis in the $y$- and $z$-directions, gives that the mean force magnitude on a contact is $f_{ij} = \sqrt{3} f_{x,ij}$. Eq. (\ref{eq:fij}) shows that the force is proportional to the area of the contact disc it forms.

Using Hertz's contact force model, we express the overlap $h$ (see Fig. \ref{fig:contact_types}a) \cite{Pop10}
\begin{align}
h = \left( \frac{3 f_{ij}}{4 E^*} \right)^{2/3} \rho^{-1/3} \ ,
\end{align}
where $\rho$ and $E^*$ were defined in the main text. Using now eq. (\ref{eq:r_overlap}) and substituting for $f_{ij}$ and $h$, we obtain
\begin{align}
r_{c, {\rm H}} = \sqrt{2} \left( \frac{6 \sqrt{3} P_o r_{c, {\rm H}}^2 \rho}{4 E^*} \right)^{1/3} \ .
\end{align}
Solving this equation for $r_{c, {\rm H}}$, we obtain finally,
\begin{align} \label{eq:rco_final}
r_{c, {\rm H}} = \frac{3 \sqrt{6} P_o} {E^*} \rho \ .
\end{align}

Note that eqs. (\ref{eq:fij}) and (\ref{eq:rco_final}) imply together $f_{ij} \propto \rho^2$. This relation can be understood on scaling grounds. Intuitively, the larger the grains, the smaller their number, and thus the heavier the load each one of them has to bear. Specifically, the number of grains scales like $N \propto \rho^{-3}$, and the external load $F$ is shared by the boundary grains, or any other planar sheet of $M \propto N^{2/3}$ grains. Putting these together we get
\begin{align} \label{eq:f_prop_r}
f_{ij} \propto \frac{F}{M} \propto N^{-2/3} \propto \rho^2 \ ,
\end{align}
in agreement with eqs. (\ref{eq:fij}) and (\ref{eq:rco_final}).

\end{appendices}

\bibliography{tpb}

\end{document}